# ATOMIC FORCE MICROSCOPY STUDY AND QUALITATIVE ANALYSIS OF MARTENSITE RELIEF IN ZIRCONIA

Sylvain DEVILLE [‡], Jérôme CHEVALIER, Hassan EL ATTAOUI

National Institute of Applied Science, Materials Department
Associate Research Unit 5510
Bat B. Pascal, 20 avenue Albert Einstein, 69621 Villeurbanne Cedex

Abstract

A recent report [S. Deville *et al.*, J. Am. Ceram. Soc., 86(12), 2225 (2003)] has shown the new possibilities offered by atomic force microscopy (AFM) to investigate martensitic transformation induced relief in zirconia. In this paper, we studied qualitatively the surface relief resulting from martensitic tetragonal to monoclinic phase transformation in yttria and ceria-doped zirconia by AFM. AFM appears as a very powerful tool to investigate martensite relief with a great precision. The phenomenological theory of martensitic crystallography could be succesfully applied to explain all the observed features. The formation conditions of the martensite are discussed, as well as ways of accommodating locally the transformation strain, i.e. self-accommodating variant pairs and microcracking. Variants growth sequences are observed. These observations bring new insights and explanations on the transformation initiation and propagation sequences.

## I. INTRODUCTION

The martensite transformation model, developed by Bain[1], has now been the object of almost a century of investigations. Its relevance to different types of materials, running from metals to ceramics, has drawn a lot of attention and studies. If the characteristics and macroscopic features of the transformation are now well predicted and understood, the number of quantitative reports of surface relief changes resulting from phase martensitic transformation is still limited, which is not surprising considering the scale at which the transformation is occurring. Martensite relief has been investigated mainly by optical methods and scanning electron microscopy, though both methods provide a limited spatial resolution, and 3D quantitative informations are not accessible. Quite fortunately, the recent development of scanning tunneling microscopy (STM) and atomic force microscopy (AFM)[2] provides the scientific community with powerful tools to investigate phenomenon's characterized by relief variations at a nanometer scale. Great progresses were accomplished in the last few years and a few reports might be found on steel based materials[3,4,5]. The absence of specific sample preparation and the possibility of observing bulk non-conductive materials make it very attractive to study martensitic transformation in ceramics in particular.

The martensitic transformation of zirconia corresponds to the tetragonal to monoclinic phase transformation[6,7]. Zirconia, when doped with yttria ($Y_2O_3$) or ceria ($CeO_2$), is indeed retained in its metastable tetragonal structure after sintering. Upon the action of mechanical stresses or hydrothermal solicitations[8-10] (i.e. water vapor at 140°C), zirconia might transform to its stable monoclinic structure. This transformation, at the origin of the transformation toughening effect[11,12], has been the object of extensive studies over the last twenty years, and its martensitic nature is now widely recognized[13]. Evidences of martensitic features were provided mainly by transmission electron microscopy and optical interferometry[14]. However, very few quantitative reports[15-17] of the martensitic features might be found in the literature due to experimental difficulties, and less interest attached to these materials so far. Recent progresses[18] in imaging resolution have shown that AFM could provide with very precise measurements of martensitic surface relief on zirconia samples. In this paper, AFM has been used to characterize surface relief resulting from martensitic transformation in ceria and yttria-doped zirconia, at a scale that was never reached before. These qualitative observations brought new insights on the transformation initiation and propagation mechanisms.

## II. EXPERIMENTAL METHODS

Ceria stabilized zirconia (Ce-TZP) materials were processed by classical powder mixing processing route, using Zirconia Sales Ltd powders, with uniaxial pressing and sintering at 1550°C for two hours. Yttria stabilized zirconia (Y-TZP) samples were processed using (3 mol.% $Y_2O_3$)-TZP powders (Tosoh, Japan), and also uniaxial pressing and sintering for two hours at 1500°C. Residual porosity was negligible. Samples were polished with standard diamond based products. Some samples were thermally etched for 12 min at 1350°C, in order to form grain boundary thermal grooves. The effect of a slight thermal etching on the ageing behaviour has been investigated, and it was shown it did not modify either the transformation mechanism or its kinetic. Thermal etching was performed to study the location of the transformation with respect to grain boundaries. Experiments were performed using both thermally etched and unetched samples.

AFM experiments were carried out with a D3100 nanoscope from Digital Instruments Inc., using oxide sharpened silicon nitride probes in contact mode, with an average scanning speed of 5 $\mu m.s^{-1}$. Since the t-m phase transformation is accompanied by a large strain (4 % vol. and 16 % shear), surface relief is modified by the formation of monoclinic phase. The vertical resolution (down to a few tenths of nanometers) of AFM allows following very precisely the transformation.

Two types of images have been obtained from AFM experiments. The first one is the height image (see Fig. 1,2 or 10), where the height of every point of the scanned surface is measured. This allows relief 3D imaging, making the image analysis and interpretation easy. The second type is the so-called derived image (see Fig. 6, 11 or 12), where the contrast is originating from the rate of relief variation, i.e. all the surfaces with the same orientation, related to the probe scanning path, will appear with the same contrast. This type of images is very convenient to discern planes with a constant angle, such as the ones forming the sides of self accommodating variant pair of martensite. Either these two types of images are presented here. The vertical and lateral scales of AFM images are always different here so as to exaggerate the relief and see it more clearly.

Ageing treatments were conducted in autoclave at 140°C, in water vapor atmosphere, with a 2 bar pressure, in order to induce phase transformation at the surface of the samples with time.

## III. RESULTS AND DISCUSSION

### A. Martensite fundamental features

The phenomenological theory of martensitic crystallography[19,20] relies on the model of an invariant plane strain (IPS) phase transformation, i.e. the habit plane of martensite does not exhibit neither strain nor rotation. Experimental evidences for this were provided here by the observation of polishing scratches at samples surface. Upon transformation, a scratch lying at the surface of transformed grains remains continuous and straight. The direction of the scratch is slightly modified, according to the overall shear of the surface, but the scratch remains unbroken. This simple observation is shown in Fig. 1. Two pairs of self accommodating martensitic variants pairs lying side by side might be observed, and scratches running across the surface are observed. It is quite clear that these scratches remained still continuous and straight, providing thus strong evidences supporting the IPS phase transformation model.

One of the most characteristic features of PTMC is the formation of self accommodating variant pairs (SAMVP). Their apparition is related in a very straightforward way to the microscopic and macroscopic effects of local shear and stresses resulting from the transformation. In the PTMC, single martensite plates might appear, leading to an N-like shape of surface relief. If two plates are growing back to back or close enough, their habit planes might join themselves, and the overall surface adopts a triangular shape. It was thought[13] the formation of SAMVP would be present only in the case were the transforming region was isolated and surrounded by untransformable material, e.g. zirconia grains in an alumina matrix, or tetragonal precipitates in MgO - partially stabilized zirconia. However, strain considerations must also be taken into account in the formation of the SAMVP. When two variants are growing back to back, their shape strain directions are opposite. Considering the very large shear strain (16%) and volume increase (4%) accompanying the t-m phase transformation, very large stresses appear in the surrounding zones of transformed material. Theses stresses might concentrate and build up to eventually stop the transformation, or they might also trigger the transformation of another neighboring system, providing certain crystallography relationships are respected. It is nevertheless worth noticing that the formation of SAMVP results in a very large reduction of long range overall shear strain, since shear in the variants of a pair are opposite and equal. This configuration is therefore very favorable from an energetical point of view. In the case of zirconia, SAMVP are present all over the surface (Fig. 1). The orientation of the pairs will be discussed later on, but it is already worth noticing this mechanism seems to be the more favorable to accommodate stresses induced by the transformation. The different ways of accommodating stresses are discussed in the next part.

### B. Martensite formation

It is worth noticing the first transformed zones present all the same relief after transformation, suggesting their orientation relationships are very similar. The junction planes of theses variants are all perpendicular to the surface (see Fig. 3, 4 or 5), so that the overall long range lateral stress is almost totally suppressed. Height variations are not restricted by the surface, so that these systems are the easiest to transform. When the transformation is propagating to the surrounding zones of the surface, due to higher stresses in the surrounding zones of the transformed regions, different crystallographical systems may be activated and transform. The relief change is consequently modified. Figure 6 provides an example of a very different orientation relationship to the surface, with two possible explanations for this. The first explanation could be related to a junction plane almost parallel to the surface, leading to a rippled surface, with much smaller height variations. However, higher residual stresses are expected in the surrounding areas in this case. The other possibility is the accommodation of deformation strain by slipping rather than twinning. It is not possible to conclude on this particular point without further local crystallographic informations.

The spatial distribution of the SAMVP at the surface is more complex. Several situations might be observed indeed. Either the SAMVP might run through the entire grain, as shown in Fig. 3, or they might also stop at the middle of the grain, and a system with a different orientation is activated, as shown in Fig. 4. Some more complex structures between these two situations might be found.

As far as the location of the variants is concerned, AFM allows the observation of very interesting features. The transformation was never initiated away from the grain boundaries, i.e. in the middle of a grain, as this would be energetically too highly unfavorable. SAMVP almost always appear at grains triple junction at first. It can be seen (Fig. 7), the transformation was indeed initiated at the triple junction before propagating to the rest of the grain. This might be interpreted by taking into account residual stress effects. It was shown[21] that residual stresses resulting from material processing concentrate at grains triple junction. As compared to other regions, these sites will act as preferential nucleation sites.

The top shape of SAMVP is of prime interest. Differences are observed between Ce-TZP and Y-TZP. In the case of Y-TZP, the junction of variant parts of a pair is always very sharp (Fig. 3 and 4). In the case of Ce-TZP, some large flat untransformed zones might be observed (Fig. 2, 5 and 9) at the junction. Since the surface of theses zones seems to be unmodified, it is reasonable to suggest they are not transformed indeed, inasmuch as this effect may be explained by the PTMC. In fact, it was shown the formation of SAMVP was a sequential process. The variants did not form all at once, and even the formation of a single variant is a sequential process. If the two variants of the pair might grow back to back, with a symmetric relief with respect to the surface, a remaining part of tetragonal phase of triangular shape is left in-between, as schematically shown in Fig. 8. When the variants are growing, stresses might add up in these untransformed zones (depending on the crystallographic relationships) until everything is transformed (Fig. 3 and 4). However, it is also possible theses stresses, if present, become so high that the transformation cannot proceed anymore[22-23]. The observed differences between Ce-TZP and Y-TZP could be explained by the differences in grain size and in crystallographic parameters.

The sequential growth of a SAMVP is illustrated in Fig. 10. The same zone of the grain was observed at two different stages of the transformation, and it might be seen the variant pair did indeed grew in height (about 10 nm) and in length. The two triangles indicate the end of the junction plane, and the distance in-between shows an increase of about 50 nm in length. This is another clear evidence that even a single pair is formed via a sequential progression.

Once some variants are formed, lots of stresses are accumulated, and the system will try to reduce its overall energy. Several strategies are possible for this. The first one is to trigger the transformation of a neighboring system, as previously mentioned. This is indeed the main mechanism observed for stress relaxation and accommodation in zirconia. However, the two systems must satisfy certain crystallographic relationships for the transformation to proceed. In the more favorable case, some coherency is found between two adjacent grains and SAMVP running from one grain to the other one might be observed, as shown in Fig. 11. Though the grain boundary thermal groove is disturbing the surface homogeneity of the pairs, the relationship between the SAMVP of the two adjacent grains is obvious. This is however a very rare occurrence in these materials, and very few transgranular martensite laths might be observed.

In a more general manner, if no specific crystallographic correspondences are found, the grain will have to accommodate all the strain when the transformation is propagating, by modifying the spatial arrangement of the SAMVP; this is particularly obvious (Fig. 5 and 12) in Ce-TZP. A system of large SAMVP is usually formed in the grain, occupying almost its entire surface. To transform the remaining parts of the grain and accommodate the strain at the same time, some smaller systems of SAMVP could be formed around, so that the resulting stresses and strains are much lower than those of larger SAMVP. A lot of these small pairs might be observed along the grain boundaries.

Finally, if no correspondence is found between two grains or two parts of a single grain, the combination of the very limited plasticity of zirconia, and the apparition of very large shear strains and stresses might lead to the formation of microcracking at the end of SAMVP. This is illustrated in Fig. 13. Thermal etching was not performed on this sample, so that there is no risk for the observed microcrack to be mistaken with a grain boundary thermal groove. The S-shaped microcrack is running across the entire micrograph. Its shape also eliminates the possibility for this artifact of being a residual polishing scratch. In all the phenomenological models developed[24,25] to explain the t-m phase transformation of zirconia, the formation of microcracks and macrocracks as a consequence of the transformation plays a major role in the propagation of the transformation. However, these cracks were never clearly observed in the surroundings zones of transformed regions. The observations reported here provide therefore strong evidence supporting these models.

### C. Relationships with the system crystallography

One of the great improvements of using AFM as compared to conventional observation methods is that it can provide 3D measurements at a nanometer scale. The lateral resolution (as low as 0.1 nm) and vertical resolution (0.01 nm) provide very reliable quantitative measurements of surface relief characteristics. For example, a precision better than 0.2° might be reached when measuring angles between planes, providing the image was acquired in good conditions, i.e. principally a probe with a radius of curvature[26] as low as possible, typically 10 nm for the best probes used here. 3D information's are nevertheless not necessary to measure angle relationships between SAMVP junction planes. In all of the observations, junction planes were always found to be either parallel or perpendicular. In regards of the original crystallographic structure (tetragonal), this is an excellent agreement with the theory, since two planes of the tetragonal cell are crystallographically equivalent, so that they might equally transform.

## IV. CONCLUSIONS

AFM has been used here for the first time to investigate precisely and qualitatively the surface characteristics of martensitic transformation in zirconia. It was shown that all the features observed here could be explained by the PTMC. The formation of SAMVP was observed. The different ways of accommodating locally the strain, i.e. SAMVP formation, system of small SAMVP formation and microcracking were observed and discussed. As far as SAMVP formation conditions are concerned, it was shown that grains triple junction appear as preferential nucleation sites.

Differences in martensitic surface relief between Ce-TZP and Y-TZP could be explained by differences in grain size.

With its unique lateral and vertical resolution, the possibility of observing bulk samples (as compared to thin foils used for transmission electron microscopy) and ease of image interpretation due to the absence of a specific interaction between the probe and the surface of zirconia, AFM appears as a unique and extremely powerful tool to investigate martensitic transformation. However, it is worth noticing no information's on the local crystallography (i.e. crystallographic orientation of the surface) might be obtained by AFM; it must be combined with different techniques. The lack of quantitative reports should vanish quite rapidly in the next few years by carrying out AFM experiments.

## ACKNOWLEDGMENTS

The authors would like to thank the CLAMS for using the nanoscope. The authors are also indebted to Pr. Guenin and Pr. Morin for their very fertile discussions on the subject.

[‡] Corresponding author: sylvain.deville@insa-lyon.fr, tel: +33 4 72 43 63 57, fax: +33 4 72 43 85 28.

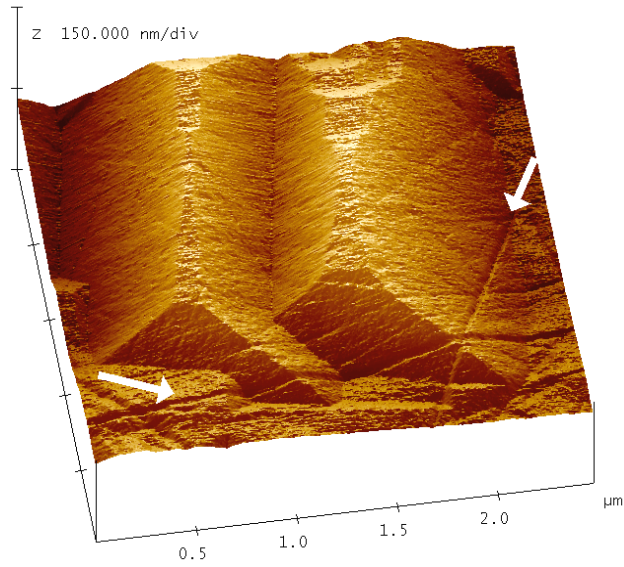

FIG. 1. Image of polishing scratches shear (arrows) after martensitic transformation in Ce-TZP.

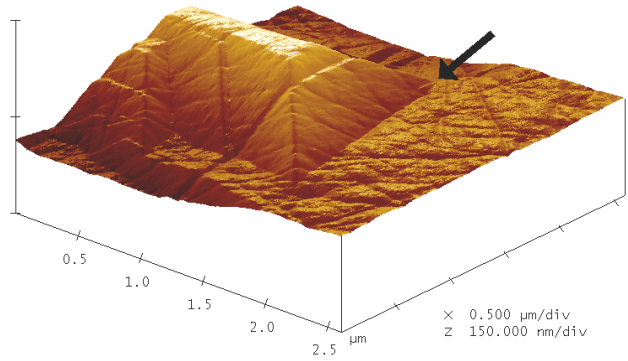

FIG. 2. Self accommodating martensitic variant pair in Ce-TZP. The intersection of the habit plane and the surface (arrow) is clearly visible.

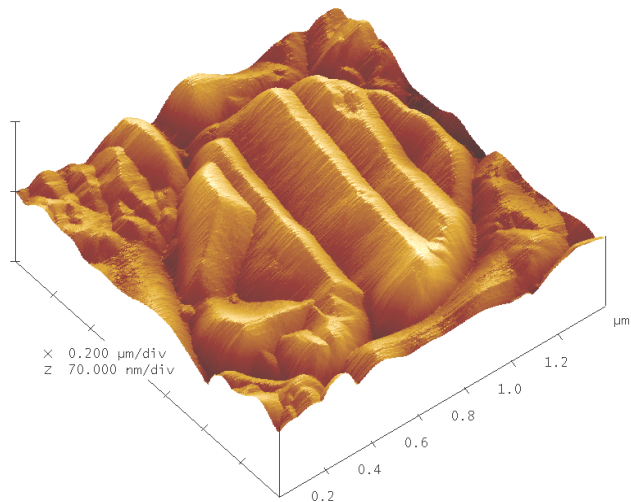

FIG. 3. Self accommodating martensitic variant pairs in 3Y-ZTP running across the entire grain.

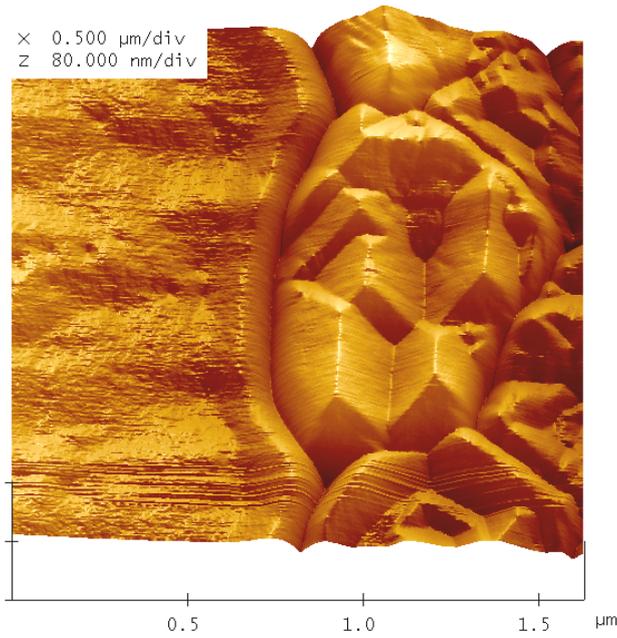

FIG. 4. Self accommodating martensitic variant pairs with a more complex spatial arrangement in 3Y-TZP. The untransformed grain on the left is a stable cubic phase grain. Grain boundary thermal grooves are clearly visible.

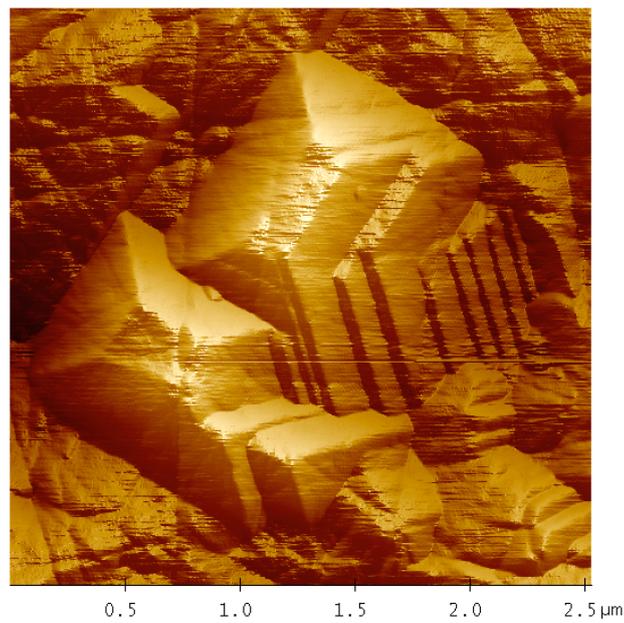

FIG. 5. Self accommodating martensitic variant pairs arrangement in Ce-TZP. Untransformed parts can be seen in-between the pairs.

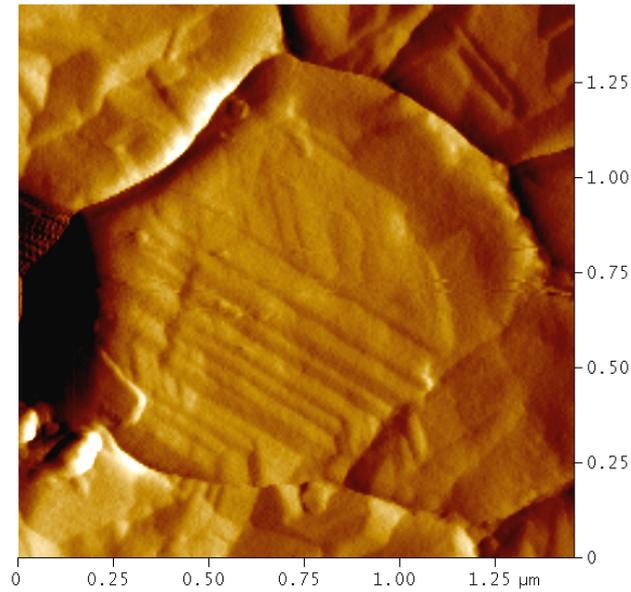

FIG. 6. Self accommodating martensitic variant pairs in 3Y-TZP with either a junction plane almost parallel to the surface, leading to a rippled surface, or with the deformation strain being accommodated by slipping rather than twinning.

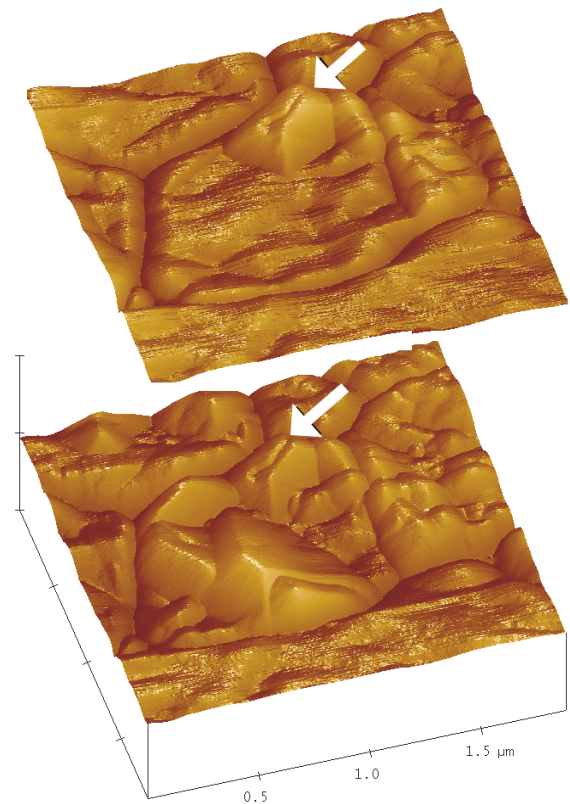

FIG. 7. Progressive transformation of a grain in 3Y-TZP. The first variants (arrow) appeared at a grain boundary triple junction.

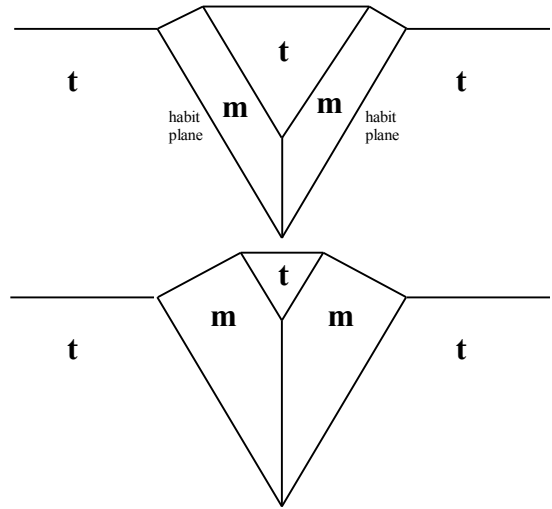

FIG. 8. Scheme of the process leading to the formation of untransformed triangular zones in Ce-TZP. t and m denotes the tetragonal and monoclinic phases.

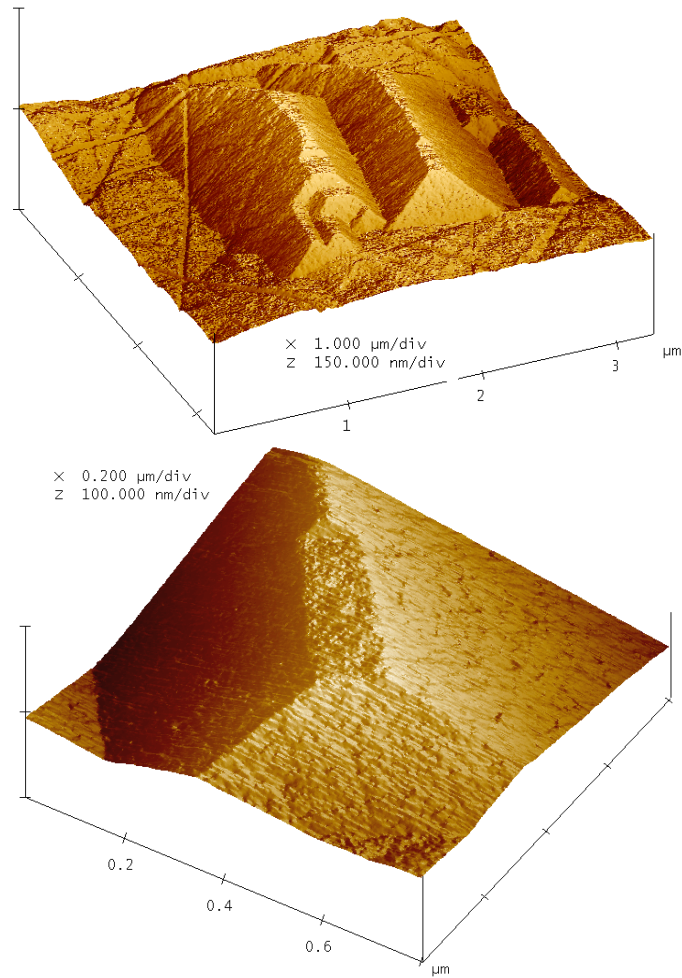

FIG. 9. Triangular untransformed zones in self accommodating martensitic variant pairs in Ce-TZP (left). A detailed zone is shown (right).

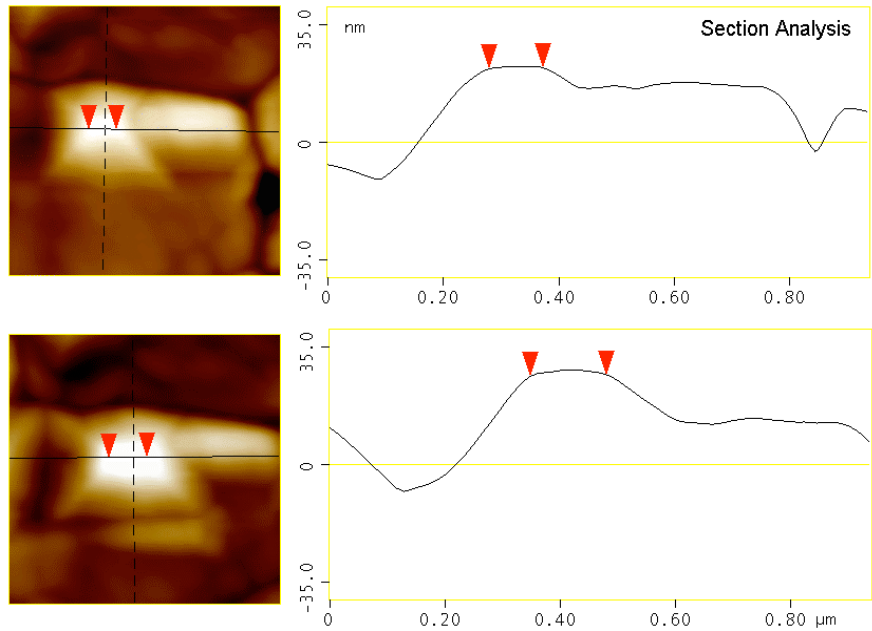

FIG. 10. Sequential growth of a variant pair in 3Y-TZP. The corresponding relief profile along the line is plotted on the right.

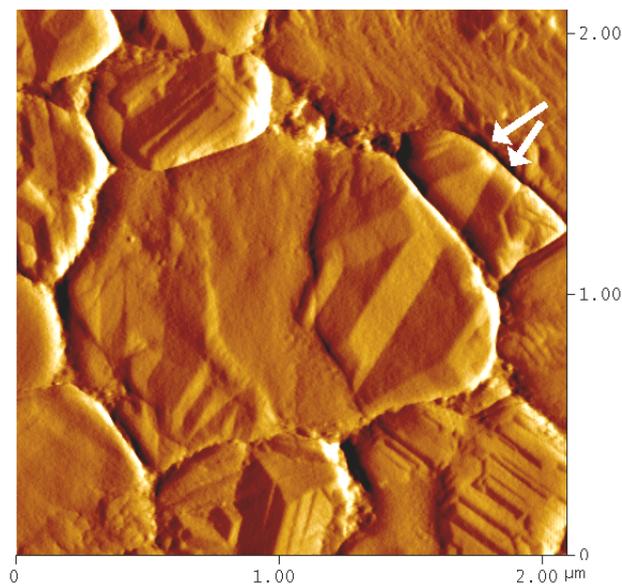

FIG. 11. Transgranular self accommodating martensitic variant pairs (arrows) in 3Y-TZP. Note the middle part of the grain has not yet transformed.

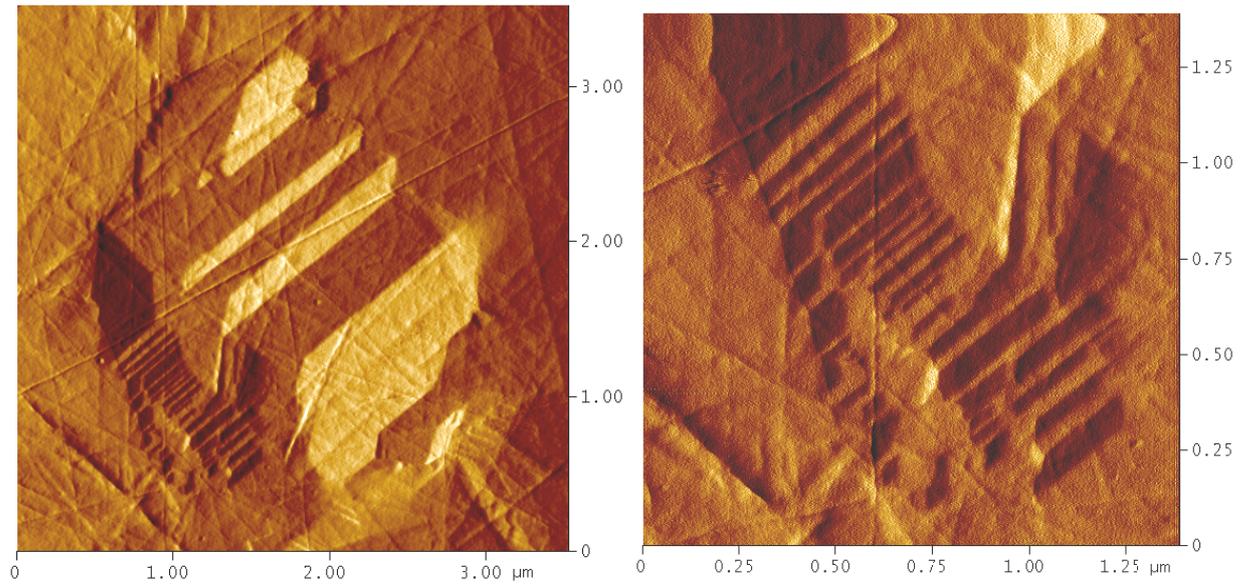

FIG. 12. Self accommodating martensitic variant pairs system in Ce-TZP, showing the formation of smaller variants along the grain boundary to accommodate locally the strain.

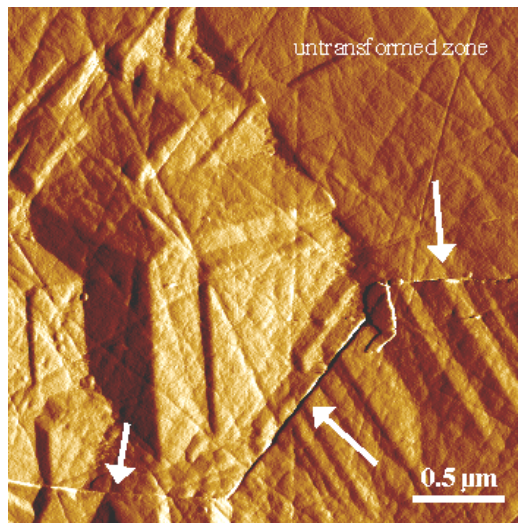

FIG. 13. Self accommodating martensitic variant pairs in Ce-TZP, showing the formation of microcracks (arrows) in the surroundings of a transformed zone.